\documentclass[prl,twocolumn,amsmath]{revtex4-1}
\usepackage{bm}
\usepackage{graphicx}
\usepackage{hyperref}
\usepackage[usenames,dvipsnames,svgnames,table]{xcolor}

\newcommand{\eps}{\varepsilon}

\newcommand{\ket}[1]{|#1\rangle}
\newcommand{\bracket}[1]{\langle #1 \rangle}

\begin{document}

\title{Optical Selection Rule of Excitons in Gapped Chiral Fermion Systems}

\author{Xiaoou Zhang}
\affiliation{Department of Physics, Carnegie Mellon University, Pittsburgh, Pennsylvania 15213, USA}

\author{Wen-Yu Shan}
\affiliation{Department of Physics, Carnegie Mellon University, Pittsburgh, Pennsylvania 15213, USA}

\author{Di Xiao}
\affiliation{Department of Physics, Carnegie Mellon University, Pittsburgh, Pennsylvania 15213, USA}

\begin{abstract}
We show that the exciton optical selection rule in gapped chiral fermion systems is governed by their winding number $w$, a topological quantity of the Bloch bands.  Specifically, in a $C_N$-invariant chiral fermion system, the angular momentum of bright exciton states is given by $w \pm 1 + nN$ with $n$ being an integer.  We demonstrate our theory by proposing two chiral fermion systems capable of hosting dark $s$-like excitons: gapped surface states of a topological crystalline insulator with $C_4$ rotational symmetry and biased $3R$-stacked MoS$_2$ bilayers.  In the latter case, we show that gating can be used to tune the $s$-like excitons from bright to dark by changing the winding number.  Our theory thus provides a pathway to electrical control of optical transitions in two-dimensional material.
\end{abstract}
\maketitle

Our understanding of optical absorption in semiconductors relies on two essential approximations~\cite{yu2010}.  The first is the effective mass approximation~\cite{luttinger1955}, in which the electron and the hole are considered as two particles moving with the effective masses of the conduction and valence bands, respectively.  In the presence of the Coulomb interaction, the electron-hole pair will form a hydrogenlike bound state known as the exciton~\cite{wannier1937}, which plays a crucial role in semiconductor optics.  The second approximation is the electric dipole approximation.  Within this approximation, the interband optical transition is usually understood in terms of the transition between atomic orbitals that make up the Bloch functions.  Together, these two approximations yield the optical selection rule for excitons, as derived in a classic paper by Elliott~\cite{elliott1957}: If the band edge transition is dipole allowed, then only the $s$-like excitons are bright and the rest are dark.  Despite its simplicity, this theory is quite versatile and can be further generalized to include complications such as band degeneracy, anisotropy, and spin-orbit interaction.

However, the validity of the above theory has been recently challenged in a new class of materials called gapped chiral fermion (CF) systems. Examples include gapped topological surface states~\cite{garate2011}, biased bilayer graphene~\cite{park2010,ju2017}, and monolayers of group-VI transition metal dichalcogenides such as MoS$_2$~\cite{xiao2012,glazov2017,wang2017}.  It has been shown that in these systems the effective mass approximation must be modified to include the Berry phase~\cite{xiao2010} carried by the CFs to give a proper account of the exciton energy spectrum~\cite{zhou2015,srivastava2015}.  At the same time, anomalous exciton optical selection rule have also been found in these systems.  For example, it has been shown that both the $s$-like and $d$-like excitons are bright in monolayer MoS$_2$, and their optical transitions have opposite circular polarization~\cite{gong2017}, while in biased bilayer graphene it is the $p$-like excitons that are bright~\cite{park2010}.  These results suggest that a new exciton optical selection rule must be established in gapped CF systems.
  
In this Letter, we show that the exciton optical selection rule in gapped CF systems  is governed by their winding number $w$ [see Eq.~\eqref{chiral} below], a topological property of the Bloch bands~\cite{fuchs2010,gao2017}.  Specifically, we find that the bright excitons in an isotropic CF system have angular momentum $m = w \pm 1$.  When the full rotational symmetry is reduced to discrete $C_N$ symmetry by crystal field effect, the allowed angular momentum of bright excitons expands to $m = w \pm 1 + nN$, where $n$ is an integer.  Our theory thus gives a unified view of the optical selection rule previously found in various gapped CF systems~\cite{garate2011,gong2017,park2010}.  To further demonstrate our theory, we propose two gapped CF systems capable of hosting dark $s$-like excitons.  The first is gapped surface states of a topological crystalline insulator with $C_4$ symmetry.  The second is $3R$-stacked MoS$_2$ bilayers.  In the latter case, we show that gating can be used to tune the $s$-like exciton from bright to dark by changing the winding number.  The value of the gate voltage to realize such a dark-bright transition is within experimental reach.  Our study, together with the previous result of the Berry phase effect on the exciton spectrum~\cite{zhou2015,srivastava2015}, provides a basic description of the electronic structure of excitons in gapped CF systems.
 
We begin with the $\bm k \cdot \bm p$ Hamiltonian for an isotropic two-dimensional CF model with an integer winding number $w$:
\begin{equation} \label{chiral}
H_0= \begin{pmatrix} \Delta & \alpha(|\bm k|)e^{iw\phi_{\bm k}} \\
\alpha(|\bm k|)e^{-iw\phi_{\bm k}} & -\Delta \end{pmatrix} \;,
\end{equation}
where $2\Delta$ is the energy gap and $\phi_{\bm k}=\tan^{-1}(k_y/k_x)$.  This Hamiltonian describes a wide range of material systems.  For example, both gapped topological surface states~\cite{garate2011} and monolayer MoS$_2$~\cite{xiao2012} have $\alpha(|\bm k|) \propto |\bm k|$ with the winding number $w = 1$, and biased bilayer graphene has $\alpha(|\bm k|) \propto |\bm k|^2$ with $w=2$~\cite{mccann2006}.  In fact, in graphene multilayers, $w$ can be made arbitrary integral values~\cite{min2008a}.  We note that this model also includes the special case of zero winding number even though it cannot be called a chiral fermion anymore.  The energy dispersion of this model is given by $\eps_{c,v} = \pm\eps_{\bm k} = \pm \sqrt{\Delta^2 + \alpha^2(|\bm k|)}$ with the corresponding eigenstates
\begin{equation}
\ket{c\bm k} = \binom{\cos\frac{\theta_k}{2}}{\sin\frac{\theta_k}{2}e^{-iw\phi_{\bm k}}} \;, \quad
\ket{v\bm k} = \binom{\sin\frac{\theta_k}{2}e^{iw\phi_{\bm k}}}{-\cos\frac{\theta_k}{2}} \;,
\end{equation} 
where $\theta_k=\cos^{-1}(\Delta/\epsilon_{\bm k})$.  The wave functions have a U(1) gauge freedom.  Here we fix the gauge by demanding that both $\ket{c\bm k}$ and $\ket{v\bm k}$ have no singularity at the band edge ($\bm k = 0$).  Under this gauge choice, the labeling of excitons by their angular momenta returns to that of the hydrogenic model in the large gap limit~\cite{zhou2015}.

An exciton in a general two-band model can be written as a linear combination of electron-hole pairs,
\begin{equation} \label{exciton}
\ket{\Psi(\bm q)} = \sum_{\bm k} f_{\bm q}(\bm k)a_{c\bm k+\bm q}^\dag a_{v\bm k}|\Omega\rangle \;.
\end{equation}
Here $|\Omega\rangle$ is the semiconductor ground state with the valence band filled and the conduction band empty, and $a_{c\bm k+\bm q}^\dag$ ($a_{v\bm k}$) creates an electron (hole) in the conduction (valence) band. The coefficient $f_{\bm q}(\bm k)$ is the exciton envelope function, where $\bm q$ and $\bm k$ are the center-of-mass and relative momentum of the electron-hole pair, respectively.  For photoexcited excitons, the center-of-mass momentum $\bm q$ is negligible, which will be set to zero and omitted hereafter.  In the isotropic model, the angular momentum $m$ is a good quantum number; thus, the envelope functions have the following form
\begin{equation}
f_m(\bm k) = \tilde{f}_m(|\bm k|)e^{im\phi_{\bm k}} \;.
\end{equation}
Finally, the oscillator strength of an exciton with angular momentum $m$ under circular polarization is given by 
\begin{equation} \label{os}
O_{m} = \frac{1}{\mu E^\text{ex}_{m}} \sum_{\eta=\pm} \Bigl|\int d\bm k\, \tilde{f}_m(|\bm k|)e^{im\phi_{\bm k}}v_\eta(\bm k)\Bigr|^2 \;,
\end{equation}
where $v_\eta(\bm k)=\langle v\bm k|\hat v_\eta|c\bm k\rangle$ is the interband matrix element of the velocity operator $\hat v_\eta =\hat v_x + i\eta\hat v_y$ with $\hat{v}_{x,y}={\partial H_0}/{\partial k_{x,y}}$, $E_m^\text{ex}$ is the exciton energy and $\mu$ is the reduced mass.

It should be pointed out that there are generally two contributions to the velocity matrix element: One is from the electron hopping between lattice sites, and the other from the dipole transition between localized orbitals~\cite{pedersen2001}.  Here we consider only the former contribution while neglecting the latter.  This is justified for the systems considered in this Letter. In MoS$_2$ the conduction and valence band edges are mainly formed by the Mo $d$ orbitals, with slight mixing from the S $p$ orbitals~\cite{liu2013}. There is no dipole transition between the even-parity $d$ orbitals, and transitions between $d$ and $p$ orbitals are negligible. Similarly, in gapped graphene systems the atomic orbitals involved are carbon $p_z$ orbitals, and optical transitions among them are dipole forbidden. 

Near the band edge, the angular dependence of the velocity matrix element is given by~\cite{supp}
\begin{equation}
\bracket{v\bm k|\hat v_\pm|c\bm k} \propto 
e^{-i({w}\mp1)\phi_{\bm k}} \;.
\end{equation}
It then follows from Eq.~\eqref{os} that after angular average only exciton states with $m=w\pm1$ have nonzero oscillator strength.  In addition, optical transitions to these two angular momentum states always have opposite circular polarization.  We emphasize that it is the $\bm k$-space phase winding of the velocity matrix element, a feature not available in the atomic transition picture, that determines the exciton optical selection rule of gapped CF systems.


Although both $w + 1$ and $w - 1$ states are bright, their oscillator strength can be quite different. For simplicity, we assume $\alpha(|\bm k|)=\alpha|\bm k|^w$. The velocity matrix elements take the form
\begin{equation}
\label{v_simple}
\begin{aligned}
\langle v\bm k|\hat v_+|c\bm k\rangle&=-2\alpha w\cos^2\frac{\theta_k}{2}k^{{w}-1}e^{-i({w}-1)\phi_{\bm k}} \;,\\
\langle v\bm k|\hat v_-|c\bm k\rangle&=2\alpha w\sin^2\frac{\theta_k}{2}k^{{w}-1}e^{-i({w}+1)\phi_{\bm k}} \;.
\end{aligned}
\end{equation}
In the large band gap limit, i.e., $\Delta\gg\alpha k_B^w$ where $k_B$ is the inverse of the exciton Bohr radius, we have $\cos{\frac{\theta_{\bm k_B}}{2}}\gg\sin{\frac{\theta_{\bm k_B}}{2}}$.  In this case, the $m={w}- 1$ exciton states are much brighter than the $m={w}+ 1$ states. 

So far, we considered only the isotropic case.  However, in a crystalline environment the $C_\infty$ symmetry is reduced to $C_N$ by the crystal field effect, which will modify the optical selection rule. 
The modifications come from two places.  First, the exciton state with angular momentum $m$ is mixed with those with angular momentum $m+nN$,
\begin{equation}
f_m(\bm k)\rightarrow\tilde f_m(|\bm k|)e^{im\phi_{\bm k}}+\sum_{n\neq 0} c_n\tilde f_{m+nN}(|\bm k|)e^{i(m+nN)\phi_{\bm k}},
\label{env_mix}
\end{equation} 
where $n$ is an integer and $c_n$ is the coefficient for each angular momentum channel, whose form has been derived in Ref.~\cite{supp}. Second, the velocity matrix element is also expanded into a series of angular momentum channels~\cite{supp}
\begin{equation}
\langle v\bm k|\hat{v}_\pm|c\bm k\rangle=\sum\limits_n v_ne^{-i(w \mp 1 + nN)\phi_{\bm k}}.
\label{v_cn}
\end{equation}
According to Eq.~\eqref{os}, the exciton selection rule now reads
\begin{equation}
m = w \pm 1 + nN \;.
\label{selection_cn}
\end{equation}
This is a reflection of the fact that in a $C_N$ invariant system the angular momentum is defined only modulo $N$~\cite{yao2008a}.  Finally, we note that the optical transitions to the $m$ and $(m + nN)$ states have the same circular polarization.

Now we examine our theory in the two previously studied systems. The first one is monolayer MoS$_2$ with winding number $w=1$. According to our theory, the $s$- and $d$-like excitons should be bright with opposite circular polarizations when the crystal field effect is ignored, and the $s$ state should be much brighter than the $d$ state due to the relatively large band gap in MoS$_2$ ($\alpha k_B/\Delta \sim 0.1$)~\cite{xiao2012,zhou2015}.  
If we turn on the crystal field, the symmetry is reduced from $C_\infty$ to $C_3$.  In this case, the $p$-like state with $m = -1$, which is dark in the isotropic model, becomes bright and has the same polarization as the $d$-like excitons with $m = 2$.  This result agrees with the direct calculation in a recent study~\cite{gong2017}.

\begin{figure}
  \includegraphics[height=65mm,width=0.45\textwidth]{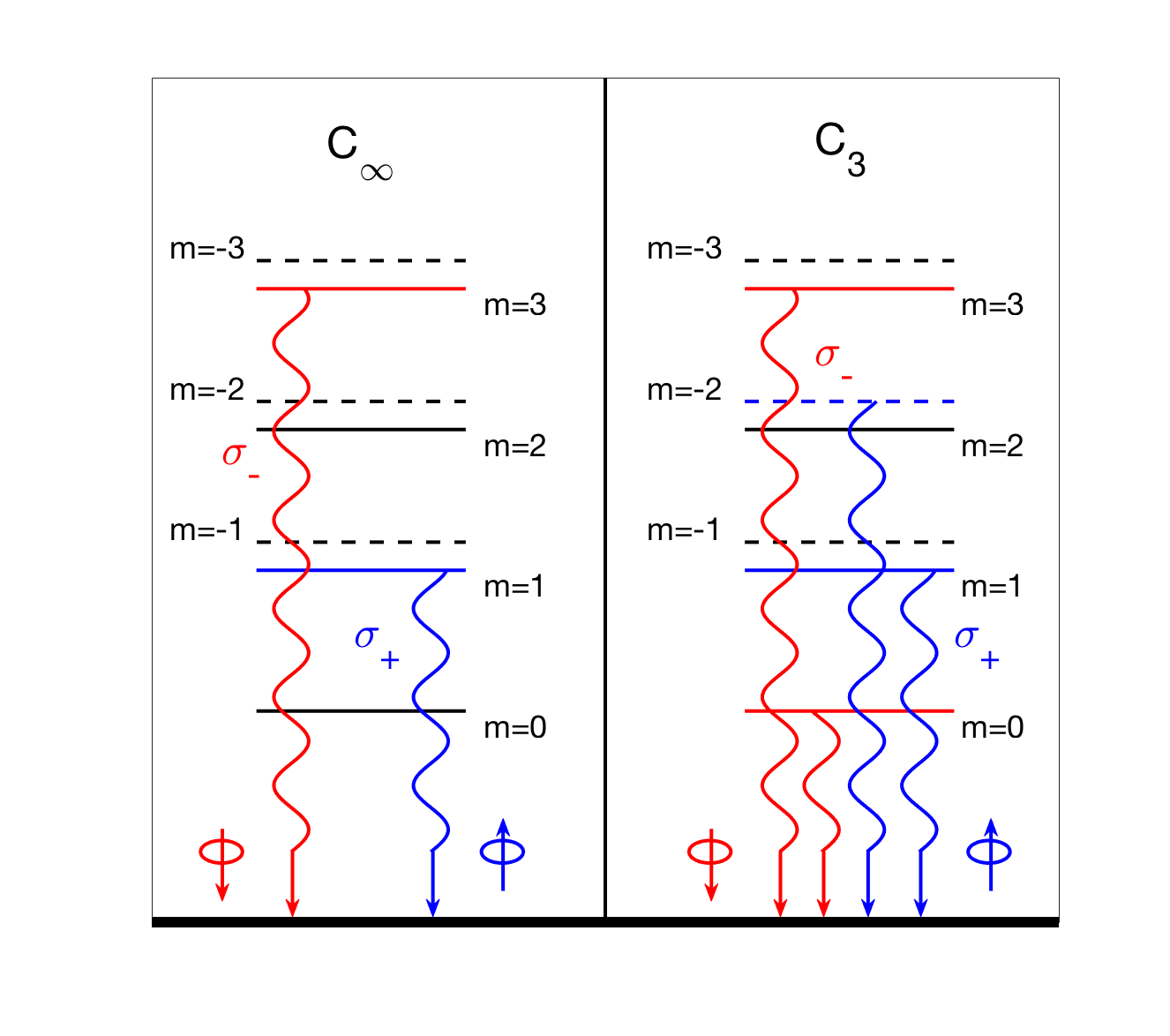}
  \caption{The exciton optical selection rule of the $w=2$ chiral fermion model when the symmetry is reduced from (a) $C_\infty$ to (b) $C_3$. The black lines indicate dark states, and the red (blue) lines are bright states with $\sigma_-$ ($\sigma_+$) polarization. The solid lines represent positive angular momenta, and the dashed lines represent negative angular momenta.}
  \label{mixing}
\end{figure}

The second example is the biased bilayer graphene~\cite{park2010}, which is described by the following effective Hamiltonian~\cite{mccann2006}
\begin{equation}
H_\text{BLG} =\begin{pmatrix} \Delta & \alpha k_+^2\\
\alpha k_-^2 & -\Delta\\ \end{pmatrix}
+ 3 \gamma_3\begin{pmatrix} 0 &  k_-\\  k_+ & 0\\
\end{pmatrix} \;,
\label{bbg_h}
\end{equation}
where $k_\pm=k_x\pm ik_y$ and $\gamma_3$ is the interlayer hopping amplitude. The first term in $H_\text{BLG}$ describes an isotropic CF model with winding number $w=2$.  This term alone would give rise to dark $s$ states, since only the $m=1$ and $m=3$ states are bright.  
However, in the presence of the $\gamma_3$ term, which reduces the $C_\infty$ symmetry to $C_3$, the optical transitions to $s$-like states are turned on and have opposite circular polarization compared to the $p$-like states. Similarly, the $m=-2$ states also become bright (see Fig.~\ref{mixing}).

To estimate the crystal field effect, we have carried out a perturbative calculation by treating $\gamma_3$ as a small quantity in the large band gap limit~\cite{supp}.  We find that the modification to the exciton envelope function is a higher-order contribution, and the main effect of the crystal field comes from its modification to the velocity matrix element, which is proportional to $\gamma_3$. Accordingly, the ratio of the oscillator strength between the $s$ and $p$ states should be proportional to $9\gamma_3^2/(2 \alpha|k_B|)^2$~\cite{supp}.  According to Ref.~\cite{park2010}, the $k$-space radius of the exciton envelope function is $k_B \sim 0.02$~\AA$^{-1}$, which gives $9\gamma_3^2/(2 \alpha|k_B|)^2 \sim 0.02$.  
Note that from a pure group theory point of view, we can also come to the conclusion that the $s$-like excitons are bright. In contrast, our theory provides a quantitative estimation of the brightness of the $s$ state.

The fact that it is the $C_3$ symmetry that turns the $s$-like excitons bright in a $w=2$ CF system suggests that by switching to a different rotational symmetry, the $s$ states can remain dark.  One such system is the gapped surface states of a topological crystalline insulator with a possible $C_4$ rotational symmetry~\cite{fu2011}.  The Hamiltonian for the surface states in such a system is given by
\begin{equation} \label{tci_zeeman}
H_{\text{TCI}}=a_1 \begin{pmatrix}
V_z & k_+^2 \\ k_-^2 & -V_z \end{pmatrix}
+ a_2\begin{pmatrix} V_z & k_-^2 \\ 
k_+^2 & -V_z \end{pmatrix} \;,
\end{equation} 
where $V_z$ is the gap opened by a time-reversal-breaking perturbation~\cite{fu2011,supp}.  We can see that this model is a mixture of CFs with $w = \pm 2$.  The simultaneous existence of both winding numbers reduces the rotational symmetry to $C_4$, and the $s$ states remain dark.

\begin{figure}
  \centering
  \includegraphics[height=32mm,width=0.2\textwidth]{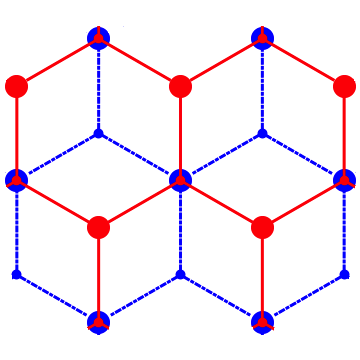}
  \caption{Top view of 3$R$-stacked MoS$_2$ bilayers. The large dots are Mo atoms and the small ones are $S$ atoms. Red (blue) dots refer to the atoms in layer $1$ ($2$).}
  \label{3r}
\end{figure}
 
\begin{figure}
  \includegraphics[height=57mm,width=0.48\textwidth]{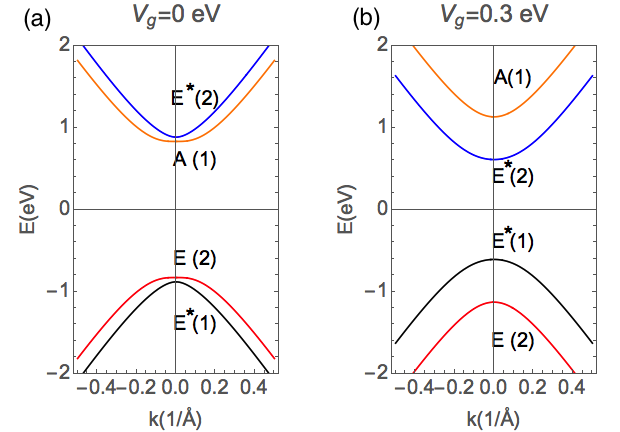}
  \caption{Band structure of a biased $3R$-MoS$_2$ bilayer at (a) $V_g = 0$ eV and (b) $V_g = 0.3$ eV.  Bands with different colors belong to different irreducible representations of $C_3$ group and layer number ($1,2$). The parameters used are $\Delta_I=0.83$ eV and $v_0=3.5$ eV \AA~\cite{Note1}.  We used a large interlayer hopping term, $\gamma_1=0.3$ eV, to make the band separation visible.}
  \label{switching}
\end{figure}

Apart from varying the symmetry group, we can also obtain dark $s$ states by switching to a different winding number while keeping the $C_3$ symmetry.  For this purpose, let us consider 3$R$-stacked MoS$_2$ bilayers.  In the $3R$-stacked bilayer structure, the top layer is shifted relative to the bottom layer along the honeycomb armchair edge, as shown in Fig.~\ref{3r}. Neglecting the spin degree of freedom, the effective Hamiltonian at one of the corners of the hexagonal Brillouin zone is given by~\cite{supp}

\begin{equation}
H_\text{3R}=\begin{pmatrix}
    \Delta_I+V_g & v_0k_- & 0 & 0\\
    v_0k_+ & -\Delta_I+V_g & \gamma_1 & 0\\
    0 & \gamma_1 & \Delta_I-V_g & v_0k_-\\
    0 & 0 & v_0k_+ & -\Delta_I-V_g\\
  \end{pmatrix} \;,
\label{3rh}
\end{equation}
where $\Delta_I$ is the gap opened by the broken inversion symmetry in each monolayer, $V_g$ is the out of plane gate voltage, and $v_0$ and $\gamma_1$ are the intralayer and interlayer hopping coefficients, respectively.  We kept only the isotropic part of the Hamiltonian, which is sufficient to demonstrate the essential physics.  An interesting feature of this system is that by varying $V_g$ one can switch the band order within the conduction and valence bands (see Fig.~\ref{switching}).  If we assume that $\Delta_I$ is large compared with the interlayer hopping constant $\gamma_1$, the critical value of the gate voltage $V_{gc}$ at the band crossing point is approximately ${\gamma_1^2}/{2\Delta_I}$.  For intralayer band gap $\Delta_I\approx0.8$ eV, and interlayer hopping $\gamma_1\approx0.05$~eV \footnote{These data are the hopping constants for 2H-stacked MoS$_2$ bilayers~\cite{wu2013}, as an estimate for 3R stacking.}, the required $V_{gc}$ is about $1.5$ meV which is not difficult to achieve in an experiment~\cite{wu2013}.  

The ability to switch the bands is important, because the winding number is a topological quantity; it can only be changed via band crossing if the rotational symmetry is kept invariant. To find the winding number before and after the band crossing we downfold the Hamiltonian~\eqref{3rh} to project out the higher conduction band and the lower valence band~\cite{winkler2003}.  Before the band crossing, i.e., $V_g<V_{gc}$, the downfolded Hamiltonian reads
\begin{equation}
H_\text{before} = \begin{pmatrix}
\Delta_I+V_g & \frac{v_0^2\gamma_1}{4\Delta_I V_g}k_-^2\\
\frac{v_0^2\gamma_1}{4\Delta_I V_g}k_+^2 & -\Delta_I-V_g\\
\end{pmatrix}\;.
\end{equation}
We can see that the winding number $w = 2$, similar to the biased bilayer graphene.  This is not surprising because each monolayer MoS$_2$ carries winding number $w = 1$, and in the 3$R$-stacking, one can simply add the winding numbers together~\cite{min2008}.  In this case, the $s$-like exciton is bright in the presence of the $C_3$ symmetry.  After the band crossing, i.e., $V_g>V_{gc}$, the $2\times2$ Hamiltonian is
\begin{equation}
H_\text{after} = \begin{pmatrix}
-\sqrt{\gamma_1^2+(\Delta_I-V_g)^2} & -\frac{v_0^2\gamma_1(\Delta_I+V_g)}{4(\Delta_I-V_g)\Delta_IV_g}k^2\\
 -\frac{v_0^2\gamma_1(\Delta_I+V_g)}{4(\Delta_I-V_g)\Delta_IV_g}k^2 & \sqrt{\gamma_1^2+(\Delta_I-V_g)^2}
\end{pmatrix}\;.
\end{equation}
Clearly the winding number is changed to $w = 0$.  Hence $m = \pm 1$ states become bright.  Turning on $C_3$ symmetry makes $m=\pm 1, \pm 4, \dots$ states bright, but the $s$ states remain dark.

Up to now, we have omitted the valley degree of freedom, which exists in most chiral fermion systems such as graphene and MoS$_2$ monolayers.  Different valleys carry an opposite winding number as a result of the time-reversal symmetry.  The corresponding optical transitions therefore have opposite circular polarization.  However, intervalley coupling of exciton states via the same circularly polarized light is unlikely since the bright exciton states in the two valleys usually have different energies (for the same circular polarization).


In conclusion, we have established a new optical selection rule of excitons in gapped CF systems.  We found that the angular momentum of bright excitons is $w \pm 1$ in the isotropic cases, and the circular polarizations of these two states are opposite.  When the crystal field effect is taken into account, the optically bright excitons have angular momentum $(w\pm1)+nN$ if the system has $N$-fold rotational symmetry.  We showed that by proper combinations of the winding number and rotational symmetry, one can engineer dark $s$ states in CF systems. The occurrence of dark excitons has already been under intense experimental investigation~\cite{ye2014,zhang2015,zhang2017,scharf2017}. Such a dark exciton has a prolonged lifetime~\cite{robert2017}, and can be utilized to realize exciton condensation~\cite{combescot2007,anankine2017,lee2017} and implement quantum information protocols~\cite{poem2010,schwartz2016}.

This work is supported by the Department of Energy, Basic Energy Sciences, Grant No. DE-SC0012509.  D.X. also acknowledges support from a Research Corporation for Science Advancement Cottrell Scholar Award.

X.Z. and W.S. contributed equally to this work.

\textit{Note added---}Upon the completion of this work, we have become aware of a recent paper, Ref.~\cite{cao2017}, which also studied the exciton optical selection rule in graphene systems.

\section*{Supplementary}

\subsection{The Velocity Matrix for the Chiral fermion model}
In the main article, we have discussed the velocity matrix in the special case $\alpha(|\bm k|)=\alpha|\bm k|^w$. In this section, we derive the velocity matrix in the more general case: $\alpha(|\bm k|)=\tilde\alpha(|\bm k|)|\bm k|^w$, where $\tilde\alpha(|\bm k|)$ is an arbitrary function of $|\bm k|$.

We start from an isotropic chiral fermion model
\begin{equation}
  H_0=
  \begin{pmatrix}
    \Delta & \tilde\alpha(|\bm k|)(k_+)^w\\
\tilde\alpha(|\bm k|)(k_-)^w & -\Delta\\
  \end{pmatrix}.
\label{hwinding_general}
\end{equation}
The velocity operator is given by $\bm v=\frac{\partial H_0}{\partial \bm k}$, such that
\begin{equation}
v_+=\begin{pmatrix}
0 & \alpha_1(|\bm k|) e^{i(w+1)\phi_{\bm k}} \\
\alpha_2(|\bm k|)e^{-i(w-1)\phi_{\bm k}} & 0\\
\end{pmatrix}\;,
\end{equation}
and
\begin{equation}
v_-=\begin{pmatrix}
0 & \alpha_2(|\bm k|)e^{i(w-1)\phi_{\bm k}}\\
\alpha_1(|\bm k|) e^{-i(w+1)\phi_{\bm k}} & 0\\
\end{pmatrix}\;,
\end{equation}
where
\begin{equation}
  \alpha_1(|\bm k|)=\frac{d\tilde\alpha(|\bm k|)}{d|k|}|\bm k|^w\;,
\end{equation}
\begin{equation}
  \alpha_2(|\bm k|)=\frac{d\tilde\alpha(|\bm k|)}{d|k|}|\bm k|^w+2w\tilde\alpha(|\bm k|)|\bm k|^{w-1}\;.
\end{equation}
In the large band gap limit, we can write the wave functions of the conduction (c) and valence band (v) to the lowest order of $|\bm k|$,
\begin{equation}
  |c\bm k\rangle=
  \begin{pmatrix}
    1\\\frac{1}{2}\frac{\tilde\alpha(|\bm k|)}{\Delta}k_-^w  \end{pmatrix}\;,\quad
  |v\bm k\rangle=
  \begin{pmatrix}
    \frac{1}{2}\frac{\tilde\alpha(|\bm k|)}{\Delta}k_+^w\\-1
  \end{pmatrix}\;.
\end{equation}
It is straightforward to show that
\begin{equation}
\label{vp}
\begin{aligned}
\langle v\bm k|v_+|c\bm k\rangle=&[\frac{\alpha^2|\bm k|^{2w}}{4\Delta^2}\alpha_1(|\bm k|)-\alpha_2(|\bm k|)]e^{i(-w+1)\phi_{\bm k}}\;,
\end{aligned}
\end{equation}
\begin{equation}
\label{vn}
\begin{aligned}
\langle v\bm k|v_-|c\bm k\rangle=&[\frac{\alpha^2|\bm k|^{2w}}{4\Delta^2}\alpha_2(|\bm k|)-\alpha_1(|\bm k|)]e^{i(-w-1)\phi_{\bm k}}\;.
\end{aligned}
\end{equation}
According to the phase winding of the velocity matrix elements, the exciton states with angular momentum $w\pm1$ always have opposite circular polarization.

\subsection{Optical Selection Rule with Crystal Field Effect}
\subsubsection{Envelope function modified by crystal field effect}\label{selection_x}
The envelope function $f_{m}(\bm k)$ for an exciton state with angular momentum $m$ is given by the Bethe-Salpeter equation \cite{zhou2015}

\begin{equation}
  \label{bs}
  \sum\limits_{\bm k'}H_{BS}(\bm k, \bm k')f_{m}(\bm k')=E_{m}^{ex}f_{m}(\bm k),
\end{equation}
where $E_{m}^{ex}$ is the exciton energy,
\begin{equation}
  H_{BS}(\bm k, \bm k')=(2\epsilon_{\bm k}+\Sigma_{\bm k})\delta_{\bm k,\bm k'}-U(\bm k,\bm k')
  \label{h_bs}
\end{equation}
is the exciton Hamiltonian, and
\begin{equation}
\label{interaction}
U(\bm k,\bm k')=V(|\bm k-\bm k'|)\langle c\bm k|c\bm k'\rangle\langle v\bm k'|v\bm k\rangle.
\end{equation}
Here $V(|\bm k-\bm k'|)$ is the Coulomb interaction in momentum space. In the isotropic case, $\epsilon_{\bm k}=\sqrt{\Delta^2+\tilde\alpha^2 k^{2{w}}}$. The self energy $\Sigma_{\bm k}$ can be absorbed into the gap and will be ignored for a qualitative discussion. 

The crystal field effect modifies ${H}_{BS}(\bm k,\bm k')$ by changing the dispersion $\epsilon_{\bm k}$ and the interaction term $U(\bm k,\bm k')$. We denote the perturbation term from the crystal field effect in the exciton Hamiltonian as ${H}_{BS}'(\bm k,\bm k')$, and the envelope function of the $s$-state ($m=0$) after perturbation is
\begin{equation}
\begin{aligned}
  f_0'(\bm k)=&f_0(\bm k)+\sum\limits_{m\neq 0}c_mf_m(\bm k)\\
  =&f_0(\bm k)\\
&+\sum\limits_{m\neq 0}\frac{\int d^2\bm k''d^2\bm k' f^*_m(\bm k''){H}'_{BS}(\bm k'',\bm k')f_0(\bm k')}{E^{ex}_0-E^{ex}_m}f_m(\bm k).
\end{aligned}
\label{env_perturb}
\end{equation}

As an application, here we consider the gapped graphene bilayer, which is described by the chiral fermion model with winding number $w=2$ in a $C_3$ invariant system. The crystal field effect gives rise to a trigonal warping term $H_{warp}$, so we have
\begin{equation}
H=H_0+H_{warp},
\end{equation}
where
\begin{equation}
  H_0=
  \begin{pmatrix}
    \Delta & \alpha(k_+)^2\\
\alpha(k_-)^2 & -\Delta\\
  \end{pmatrix}
\label{cf2}
\end{equation}
and
\begin{equation}
  H_{warp}=
  3\gamma_3\begin{pmatrix}
    0 & k_-\\
k_+ & 0\\
  \end{pmatrix}.
  \label{warp2}
\end{equation}

In this case, the perturbation term $H_{BS}'$ is dominated by the correction of the band gap, which gives
\begin{equation}
H'_{BS}=\frac{3\gamma_3\alpha}{\Delta}(k_+^2+k_-^3)\;.
\end{equation}
In the large band gap limit $\Delta\gg\tilde\alpha k_B^w$, we can estimate the coefficients $c_m$ according to Eq.~\eqref{env_perturb} as
\begin{equation}\label{coen3}
c_{-3}\propto \frac{\alpha|k_B|^2}{E^{ex}_0-E^{ex}_{-3}}\frac{3\gamma_3|k_B|}{\Delta}\;,
\end{equation}
\begin{equation}\label{coe3}
c_{3}\propto \frac{\alpha|k_B|^2}{E^{ex}_0-E^{ex}_{3}}\frac{3\gamma_3|k_B|}{\Delta}\;.
\end{equation}

All wave vectors are estimated by $k_B=2\pi/a_B$ and $a_B$ is the exciton Bohr radius. The first terms in Eq.~\eqref{coen3} and \eqref{coe3} have an order of magnitude of 1. The second terms are proportional to the warping term $\gamma_3$ which is regarded as a small quantity compared with the band gap $\Delta$. Note that in the first-order perturbation theory, $c_3$ and $c_{-3}$ are the only nonzero coefficients. With higher-order perturbation theory, $c_{3n}$ will become nonzero, where $n$ is an integer.

\subsubsection{Velocity operator modified by crystal field effect}\label{selection_v}
For the Hamiltonian Eq.~\eqref{cf2} and \eqref{warp2},  the wave functions of the conduction and valence band are
\begin{equation}
  |c\bm k\rangle=
  \begin{pmatrix}
    1\\
    \frac{\alpha}{2\Delta}k_-^2+\frac{3\gamma_3}{2\Delta}k_+
  \end{pmatrix}
\end{equation}  
\begin{equation}
  |v\bm k\rangle=
  \begin{pmatrix}
    \frac{\alpha}{2\Delta}k_+^2+\frac{3\gamma_3}{2\Delta}k_-\\-1
  \end{pmatrix}
\end{equation}
in the large band gap limit. To the leading order of $\gamma_3$, we can show that
\begin{equation}
\label{vp}
\begin{aligned}
\langle v\bm k|v_+|c\bm k\rangle=&-4\alpha |k|e^{-i\phi_{\bm k}}\\
&+\frac{\alpha^2|k|^{4}}{4\Delta^2}6\gamma_3e^{-4i\phi_{\bm k}}\;,
\end{aligned}
\end{equation}
\begin{equation}
\label{vn}
\begin{aligned}
\langle v\bm k|v_-|c\bm k\rangle=&\frac{\alpha^2|k|^{4}}{\Delta^2}\alpha |k|e^{-3i\phi_{\bm k}}\\
&+(\frac{\alpha^2|k|^4}{\Delta^2}6\gamma_3-6\gamma_3)\;.
\end{aligned}
\end{equation}

\subsubsection{Selection rule modified by crystal field effect}
Combining the results from section~\ref{selection_x} and~\ref{selection_v}, we can evaluate the oscillator strength modified by the crystal field effect.

As mentioned in the main text, the $s$-state is dark in the isotropic $w=2$ chiral fermion model, and is bright when the $C_3$ warping term is included. For a quantitative discussion, we calculate the oscillator strength for the $s$-state in gapped graphene bilayers,
 \begin{equation}\label{osc3}
\begin{aligned}
  O_{0}=&\frac{1}{\mu E^{ex}_{m}}\sum\limits_{\eta=\pm}|\sum\limits_{\bm k}f_{m}(\bm k)v_\eta(\bm k)|^2\\
  =&\sum\limits_{\eta=\pm}|\sum\limits_{\bm k}(\tilde f_0+\sum\limits_{n=\pm1}c_{3n}\tilde f_{3n}e^{3ni\phi_{\bm k}})(\sum\limits_{n'=\pm1}v^{3n'}_\eta e^{3n'i\phi_{\bm k}})|^2\\
  \approx& |\sum\limits_{|k|}6\tilde f_{0}\gamma_3|^2,
\end{aligned}
\end{equation}
where $v^{m}_\eta(|\bm k|)=\int d\phi_{\bm k}e^{-im\phi_{\bm k}}v_\eta(\bm k)$ and $f_m(\bm k)=\tilde{f}_m(|\bm k|)e^{im\phi_{\bm k}}$. In the last line of Eq.~\eqref{osc3}, only the leading order terms are preserved, which come from the modification of the velocity matrix. Note that the perturbation terms from the envelope function are high order corrections.\par
Now we compare the oscillator strength of the $s$-state with that of the bright $p$-state ($m=1$) and $m=3$ state. The oscillator strengths of $p$-state and $m=3$ state are given by
\begin{equation}
O_1=|\sum_{|k|}\tilde f_1 4\alpha |k||^2\;,
\end{equation}
\begin{equation}
O_3=|\sum_{|k|}\tilde f_3 \frac{\alpha^2|k|^{4}}{\Delta^2}\alpha |k||^2\;.
\end{equation}
The relative oscillator strength can be written as
\begin{equation}
O_0/O_1\sim (\frac{3\gamma_3}{2\alpha |k_B|})^2
\end{equation}
 and 
\begin{equation}
O_0/O_3\sim(\frac{4\Delta^2}{\alpha^2|k_B|^{4}})^2O_0/O_1.
 \end{equation}
 Accordingly, the relative oscillator strength between the $s$-state and the $p$-state is given by the ratio between the off-diagonal terms in $H_{warp}$ and $H_0$, which are ${\gamma_3|k_B|}$ and ${\alpha |k_B|^2}$, respectively. Moreover, compared with $O_0/O_1$, the relative oscillator strength $O_0/O_3$ is enhanced by a factor of $\frac{4\Delta^2}{\alpha^2|k_B|^{4}}$ , which is large in the large band gap limit.\par
 
\subsection{The Gapped Surface State of the $C_4$ invariant Topological Crystalline Insulator}
According to Fu's paper~\cite{fu2011}, the effective Hamiltonian of the  topological crystalline insulator surface state with $C_4$ symmetry is given by
\begin{equation}
  H_s=\frac{k^2}{2m_0}I_{2\times2}+\frac{k_x^2-k_y^2}{2m_1}\sigma_z+\frac{k_xk_y}{2m_2}\sigma_x\;.
  \label{tci_0}
\end{equation}
Note that this Hamiltonian is written in the basis of $p_x$ and $p_y$ orbital. We can transform the basis from $p_{x,y}$ into $p_\pm$ by applying a unitary transformation
\begin{equation}
  U=\frac{1}{\sqrt{2}}
  \begin{pmatrix}
    1&1\\i&-i
  \end{pmatrix}.
\end{equation}
Ignoring the $I_{2\times2}$ term, the effective Hamiltonian~\eqref{tci_0} can be conveniently written in the following form
\begin{equation}
   \label{tci_zeeman}
\begin{aligned}
   H'_s=&a_1
\begin{pmatrix}
  V_z & k_+^2\\
  k_-^2 & -V_z\\
\end{pmatrix}+
a_2\begin{pmatrix}
  V_z & k_-^2\\
  k_+^2 & -V_z\\
\end{pmatrix}.
\end{aligned}
 \end{equation}
This model is a mixture of chiral fermions with winding number $\pm2$, which gives Eq.~(12) in the main text. The weight for this mixing is given by $a_1$ and $a_2$.
\subsection{Effective Model for 3R Stacking MoS$_2$ Bilayer}
\subsubsection{The irreducible representations of the basis functions}
The effective four-band model of the $3R$-stacked MoS$_2$ bilayer can be derived from a pure symmetry analysis. The first step is figuring out the irreducible representations of the four basis, which can be seen by deriving the transformation law of the atomic wave functions and the plane wave part of Bloch wave functions under the symmetry operations. On one hand, the band edge states consist of $d_{z^2}$ and $d_{x^2-y^2}+id_{xy}$ orbitals on Mo atoms from both layers at valley $K=(-\frac{4\pi}{3\sqrt{3}a},0)$, where $a$ is the lattice constant of MoS$_2$. The atomic orbital  $d_{z^2}$ is invariant under a clockwise three-fold rotation $\hat C_3$ while $\phi(r)=d_{x^2-y^2}+id_{xy}$ orbital gets a phase factor $e^{-i\frac{2\pi}{3}}$. On the other hand, the plane wave part of the Bloch function also gives rise to a phase factor under a three-fold rotation. Together, the transformation rule follows
\begin{equation}
  \begin{aligned}
   C_3 \psi_{cK}(r)=&\sum\limits_{\bm R_i}e^{i\bm K\cdot (\bm R_i+\bm{\tilde{R}})}C_3\phi(\bm r-(\bm R_i+\bm{\tilde{R}}))\\
=&\sum\limits_{\bm R_i}e^{i\bm K\cdot(\bm R_i+\bm{\tilde{R}})}\phi(C_3^{-1}\bm r-(\bm R_i+\bm{\tilde{R}}))\\
=&\sum\limits_{\bm R_i}e^{i\bm K\cdot(\bm R_i+\bm{\tilde{R}})}\phi(\bm r-C_3(\bm R_i+\bm{\tilde{R}}))e^{i\alpha}\\
=&e^{i(C_3\bm K-\bm K)\cdot\bm{\tilde{R}}}e^{i\alpha}\psi_{cK}(r).\\
  \end{aligned}
\end{equation}
Here $\bm R_i$ is the position of the unit cell, $\bm{\tilde{R}}$ is the relative position of the lattice point in the unit cell, and $\alpha$ is the phase contributed by the atomic orbital under a three-fold rotation. As mentioned earlier, we have $\alpha=0$ for $d_{z^2}$ and $\alpha=-\frac{2\pi}{3}$ for $d_{x^2-y^2}+id_{xy}$. According to the top view of the 3R-stacked MoS$_2$ bilayer in Fig.~\ref{3r}, the $R_A$ point is the rotation center of the three-fold rotation, so $\bm R_A=0$, and $\bm R_B=(\frac{\sqrt{3}}{2}a,\frac{1}{2}a)$. Combined with $C_3\bm K-\bm K=(\frac{2\sqrt{3}\pi}{3a},\frac{2\pi}{3a})$, and $(C_3\bm K-\bm K)\cdot \bm R_B=-\frac{2\pi}{3}$, the $d_{z^2}$ and $d_{x^2-y^2}+id_{xy}$ bands on layer $1$ (red in Fig.~\ref{3r}) are in $A(1)$ and $E^*(1)$ irreducible representations, while on layer $2$ (blue in Fig.~\ref{3r}) they are in $E^*(2)$ and $E(2)$ irreducible representations. The numbers in the brackets label the layers where the states are located in.
\subsubsection{Effective $k\cdot p$ Hamiltonian} 
Using $\bm k\cdot \bm p$ expansion, the matrix element between two basis $\ket{a}$ and $\ket{b}$ is proportional to $\bm k\cdot\bracket{a|\bm{\hat p}|b}$, where $\bm{\hat p}$ is the momentum operator. In order to have a non-vanishing $\langle b|\hat{p}_\eta|a \rangle$, the direct product of the irreducible representations $(\Gamma^b)^*\bigotimes\Gamma^{p_\eta}\bigotimes\Gamma^a$ should contain the unit representation. Here the index $\eta$ represents the polarization of the momentum operator.
 For example, $\langle A(1)|\hat{p}_+|E^*(1) \rangle$ is nonzero, since $\Gamma^{p_+}$ is in $E$ representation and $A\bigotimes E\bigotimes E^*$ gives unit representation. As a consequence, the matrix element between $A(1)$ and $E^*(1)$ basis should be proportional to $k_x-ik_y$. Following this argument, the effective four-band Hamiltonian written in the basis of  $A(1)$, $E^*(1)$, $E^*(2)$, $E(2)$ is given by
\begin{equation}
\label{3rh4}
\begin{aligned}
  &H_\text{3R}=\\
  &\begin{pmatrix}
    \Delta_I & v_0(k_x-ik_y) & v_1(k_x-ik_y) & v_3(k_x+ik_y)\\
    v_0(k_x+ik_y) & -\Delta_I & \gamma_1 & v_2(k_x-ik_y)\\
    v_1(k_x+ik_y) & \gamma_1 & \Delta_I & v_0(k_x-ik_y)\\
    v_3(k_x-ik_y) & v_2(k_x+ik_y) & v_0(k_x+ik_y) & -\Delta_I\\
  \end{pmatrix}.
\end{aligned}
\end{equation}
\begin{figure}
  \centering
  \includegraphics[height=32mm,width=0.2\textwidth]{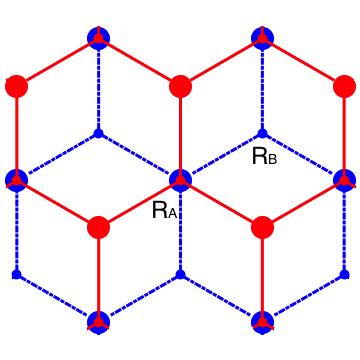}
  \caption{Top view of 3$R$-stacked MoS$_2$ bilayers. The large dots are Mo atoms and the small ones are $S$ atoms. Red (blue) dots refer to the atoms in layer $1$ ($2$). The lattice constant is $a$.}
  \label{3r}
\end{figure}
The diagonal $2\times2$ blocks are the Hamiltonians for the two MoS$_2$ monolayers, and the off-diagonal blocks represent the interlayer coupling. Assuming that the interlayer coupling is weak, we can ignore the $v_1$, $v_2$ and $v_3$ terms which are zero at $\bm k=0$.

\subsubsection{Downfolding into the two-band model}
In the simplified case $v_1=v_2=v_3=0$ and by applying an out-of-plane gate voltage $V_g$, we have the four-band model of the MoS$_2$ bilayer
\begin{equation}\label{3rhnew}
\begin{aligned}
  &H_\text{3R}=\\
  &\begin{pmatrix}
    \Delta_I+V_g & v_0 k_- & 0 & 0\\
    v_0 k_+ & -\Delta_I+V_g & \gamma_1 & 0\\
    0 & \gamma_1 & \Delta_I-V_g & v_0 k_-\\
    0 & 0 & v_0 k_+ & -\Delta_I-V_g\\
  \end{pmatrix}\;.
\end{aligned}
\end{equation}
This produces Eq.~(13) in the main text. Note that the $C_\infty$ symmetry is retained in this simplified model.

 We first diagonalize the four-band model~\eqref{3rhnew} at $\bm k=0$ by a unitary transformation $U$. To the first order of $\gamma_1/\Delta_I$,
\begin{equation}
U=\begin{pmatrix}
1 & 0 & 0 & 0\\
0 & 0 & 1 &\frac{\gamma_1}{2E_4}\\
0 & 0 & -\frac{\gamma_1}{2E_4} & 1\\
0 & 1 & 0 & 0\\
\end{pmatrix}\;,
\end{equation}
where $E_4=\sqrt{(\Delta_I-V_g)^2+\gamma_1^2}$. Applying the unitary transformation $U$ on $H_\text{3R}$, we have
\begin{equation}
\begin{aligned}
\tilde H_\text{3R}&=U^{-1}H_\text{3R}U\\
=&\begin{pmatrix}
    \Delta_I+V_g & 0 & v_0k_- & \frac{\gamma_1v_0}{2E_3}k_-\\
    0 & -\Delta_I-V_g & -\frac{\gamma_1v_0}{2E_3}k_+ & v_0k_+\\
    v_0k_+ & -\frac{\gamma_1v_0}{2E_3}k_- & -E_4 & 0\\
    \frac{\gamma_1v_0}{2E_3}k_+ & v_0k_- & 0 & E_4\\
  \end{pmatrix}\;.
\end{aligned}
\end{equation}

Now the basis are in $A(1)$, $E(2)$, $E^*(1)$, $E^*(2)$ irreducible representations. When $V_g$ is small, the lowest conduction band and the highest valence band are made up of the upper two basis $A(1)$, $E(2)$. In this case, we downfold the four-band model into a two-band model that only includes the $A(1)$ and $E(2)$ basis. To the second order of $\bm k$, the downfolded Hamiltonian is given by~\cite{winkler2003}
\begin{equation}
\begin{aligned}
(H_\text{before})_{mm'}&=(\tilde H_\text{3R}^0)_{mm'}\\
+&\sum_l\frac{(\tilde H_{\text{3R}}^1)_{ml}(\tilde H_{\text{3R}}^1)_{lm'}}{2}(\frac{1}{E_m-E_l}+\frac{1}{E_{m'}-E_l}),
\end{aligned}
\end{equation}
where $m,m'=1,2$, $l=3,4$, $E_n=(\tilde H_{\text{3R}})_{nn} $ and $\tilde H_{\text{3R}}^0$, $\tilde H_{\text{3R}}^1$ contain the diagonal and off-diagonal terms in $\tilde H_{\text{3R}}$, respectively. It is straightforward to get the downfolded Hamiltonian
\begin{equation}
H_\text{before} = \begin{pmatrix}
\Delta_I+V_g & \frac{v_0^2\gamma_1}{4\Delta_I V_g}k_-^2\\
\frac{v_0^2\gamma_1}{4\Delta_I V_g}k_+^2 & -\Delta_I-V_g\\
\end{pmatrix}\;.
\end{equation}
Here we get a chiral fermion model with $w=2$, as given by Eq.~14 in the main text. This model is similar to biased graphene bilayers where the $s$-state is bright in a $C_3$ invariant system.

When $V_g$ is large enough to switch the band order, the lowest conduction band and the highest valence band are given by $E^*(1)$ and $E^*(2)$ basis. Similar to $H_\text{before}$, we get the downfolded two-band model after the band crossing
\begin{equation}
H_\text{after} = \begin{pmatrix}
-\sqrt{\gamma_1^2+(\Delta_I-V_g)^2} & -\frac{v_0^2\gamma_1(\Delta_I+V_g)}{4(\Delta_I-V_g)\Delta_IV_g}k^2\\
 -\frac{v_0^2\gamma_1(\Delta_I+V_g)}{4(\Delta_I-V_g)\Delta_IV_g}k^2 & \sqrt{\gamma_1^2+(\Delta_I-V_g)^2}
\end{pmatrix}\;.
\end{equation}
We see that the winding number is changed from $2$ to $0$ after the band crossing. In this case, the $s$-state is always dark even when the discrete $C_3$ symmetry is considered.


%

\end{document}